\documentclass[12pt]{article}
\usepackage{epsfig}
\begin{document}

\title{\bf A ``Freely Coasting'' Universe }

\author{Savita Gehlaut, A. Mukherjee, S. Mahajan \& D. Lohiya\thanks{\small 
Inter University Centre for  Astronomy and Astrophysics,
Postbag 4, Ganeshkhind, Pune 411 007, India; dlohiya@iucaa.ernet.in}}

\date{}
\maketitle 

\vspace{-1.2cm}
\begin{center}
\em {Department of Physics and Astrophysics, University of Delhi, 
Delhi--110 007, India} 
\end{center}

\begin{abstract}

	A strictly linear evolution of the cosmological scale factor
is surprisingly an excellent fit to a host of cosmological observations.
Any model that can support such a coasting presents itself as a 
falsifiable model as far as classical cosmological tests are concerned.
This article discusses the concordance of such an evolution in relation
to several standard observations. Such evolution is known to be
comfortably concordant with the Hubble diagram as deduced from 
current supernovae 1a data, it passes constraints arising from the age and 
gravitational lensing statistics and just about clears 
basic constraints on nucleosynthesis. Such an evolution exhibits 
distinguishable and verifiable features for the recombination era. 
The overall viability of such models is discussed.

\end{abstract}

\pagebreak
 
\vspace{.5cm} 
\section 
{\bf 1. INTRODUCTION}
\vspace{.5cm}

	Large scale homogeneity and isotropy of matter and radiation
observed in the universe
suggests the following [Friedmann-Robertson-Walker (FRW)] form
for the space-time metric:
\begin{equation}
\label{1}
ds^2 = dt^2 - a(t)^2[{dr^2\over {1 - Kr^2}} + r^2(d\theta^2 
+ sin^2\theta d\phi^2)]
\end{equation}
Here $K = \pm 1, 0$ is the curvature constant.
In standard ``big-bang'' cosmology, the scale factor $a(t)$ is completely
determined by the model for the equation of state of matter  
and Einstein's equations. 
The scale factor, in turn, determines the response of a chosen 
model to cosmological observations. Four decades ago, the main ``classical'' 
cosmological tests were (1) The galaxy number count as a function of 
red-shift; (2) The angular diameter
of ``standard'' objects (galaxies) as a function of red-shift; and finally
(3) The apparent luminosity of a ``standard candle'' as a function of
red-shift.   Over the last two decades, other  tests 
that have been perfected, or are fast approaching the state of perfection, 
are: the early universe nucleosynthesis constraints, 
estimates of age of the universe in comparison to ages of old objects,
statistics of gravitational lensing and finally, the physics of recombination
as deduced from cosmic microwave background anisotropy.

	In this article we explore concordance of the above observations 
with a FRW cosmology in which the scale factor evolves linearly with time:
$a(t) \propto t$, right from the creation event itself. 
The motivation for such an endeavor comes from several considerations.
First of all, such a cosmology does not suffer from the horizon problem. 
Horizons occur in models with $a(t) \approx t^\alpha$ for $\alpha < 1$
[see eg. \cite{kolb,borner}]. 
As a matter of fact, a linearly evolving model is the only power law model
that has neither a particle horizon nor a cosmological event horizon.
Secondly,  linear evolution of the scale factor is supported in alternative 
gravity theories where it turns out to be independent of the matter 
equation of state \cite{meetu,dol,ford}. The scale factor in such theories
does not constrain the matter density parameter. This contrasts with
the Standard FRW model where the Hubble parameter determines a critical
value of density which turns out to be a dynamical repeller. This is
the root cause of the ``flatness'' or fine tuning  problem.  Finally, such 
a linear coasting cosmology, independent of the equation of state of matter, 
is a generic feature in a class of models that attempt to dynamically 
solve the cosmological constant
problem \cite{wein,dol,ford}. Such models have a scalar 
field non-minimally coupled
to the large scale scalar curvature of the universe. 
With the evolution of time, the non-minimal coupling
diverges, the scale factor quickly approaches linearity and the non-minimally
coupled field acquires a stress energy that cancels the vacuum energy 
in the theory.  

	There have been other gravity models that also account for a linear
evolution of the scale factor. Notable among such models is
Allen's \cite{allen} in which such a scaling 
results in an $SU(2)$ cosmological 
instanton dominated  universe. Yet another possibility 
arises from the Weyl gravity theory of Manheim and 
Kazanas \cite{mann}. Here again the FRW scale factor
approaches a linear evolution at late times. 

	Although any of the above are good enough reasons for exploring 
the concordance of a linear coasting, we add to this list the following 
reason of our own. The averaging problem in General Relativity has never 
been properly addressed, let alone solved \cite{ellis,ehlers}. This
is in contrast with
 the corresponding problem in classical electromagnetic theory
\cite{jackson}.
There one can (i) start with multi-singular solutions to the Laplace equation,
(ii) smear each charge over a large enough sphere, and (iii) if the
overall distribution satisfies Dirichlet / Neumann boundary conditions at
infinity,
the average potential can be defined and coincides with the solution to the 
Poisson equation. In General Relativity the corresponding construction
has not been carried out. All precision
tests of General Relativity strictly involve vacuum (source free region) 
solutions of Einstein 
equations. Strictly speaking, there are no tests of Einstein theory
with matter. In the interior of all astrophysical sources, 
either the weak field 
(Newtonian) limit is put to test or, where the weak field limit is expected to
break down, one assumes General Relativity to parametrize the equation of 
state (eg. for neutron / quark stars etc.). 

	On the other hand, the above 
problems could be circumvented by taking Einsteins equations with the 
source terms as the {\it defining} equations for a gravity theory. The 
justification for such an approach could rely on its correct Newtonian
limit. Such an attitude comes with its own problems. 

	 First of all, one encounters a related {\it averaging}
problem again when one applies the theory to cosmology. 
Is it justified to assume that the large scale behavior of the 
lumpy universe to be the same as that predicted by the smoothed
out FRW models ? The essential issue is that  averaging the metric
does not commute with determining the local connection followed by
the determination of the local Ricci 
tensor and finally forming the field equations to determine the metric. 
There have been several
attempts to resolve this issue \cite{ellis,ehlers}, but with limited
success. Moreover, reliance on an ansatz just because of its Newtonian
limit may in fact be flawed. Newtonian gravity does not offer unique 
cosmological solutions in the continuum limit for an open cosmology
\cite{buch1997}. 

	All studies on the averaging problem and the 
continuum limit have not considered the retarded effects in their full 
generality. Newtonian cosmology, applied to an exploding {\it Milne ball}
in a flat space-time [see eg. 
\cite{milne,rindler}] gives a unique linear coasting cosmology viz. the FRW 
[Milne] metric with $a(t) = t$. 

	Finally, we recall an approach to General Relativity starting from
a spin two field interacting with a source in a flat space-time.
Incorporating back reaction on the source in a gauge invariant manner and 
to all orders of perturbations
yields Einstein's theory \cite{Deser,Krai,Feyn,Gupta}. 
However, 
the entire analysis relies on canonical propagation of gravity and fails
for a distribution of particles across horizons if one has a cosmological
creation event. Equivalence Principle tells us that the  natural 
way to describe a distribution of particles 
just after a creation event, in case one demands gravity not to have globally
set in
on account of event horizons, is a distribution in a flat space-time. This 
again takes one back to Milne's cosmology.

	Indeed, consider the universe just after its ``creation event'',
defined  at $t=0$, at a small enough time $t = \epsilon$ after its creation.
In a classical description, let the matter be distributed as a swarm of 
particles in a Reimannian manifold. One may accept Einstein'e theory as a 
local theory and invoke Einstein's equations at the location of each 
particle, viz.: 
$G_{\mu\nu} = -8\pi T_{\mu\nu}$. In the inter-particle spaces, the equations 
read: $G_{\mu\nu} = 0$. For $\epsilon$ small enough, there is no reason to 
expect the global space-time dynamics to be governed by an average stress 
energy distribution: $<G_{\mu\nu}> = -8\pi <T_{\mu\nu}>$. 
This is particularly unreasonable on account of horizons in the theory. There
is absolutely no dynamical reason to expect an {\it average} gravity,
 described by
Einstein's equations on the average, to have globally ``set in''. It is much
more reasonable to expect gravity {\it not} to have set in globally on account
of {\it retarded effects}. Global matter distribution on large
scale, in the absence of global gravitation  set in, is naturally 
described  as a distribution in a flat space-time. Such a general
homogeneous and isotropic distribution of matter in a flat space-time, 
described in Co moving coordinates, is just the Milne ball. This reduces to an
open FRW universe with the scale factor $a(t) = t$.

	 We may take any of the above as the basis for our linear coasting conjecture. 
In what follows, we assume that an homogeneous background FRW universe
is born and evolves as a Milne Universe about which a matter distribution
and standard General
Relativity would determine the growth of perturbations. Thus we conjecture
that Einstein equations give a correct {\it microscopic} description of 
gravitation. This being so, the global dynamics of a FRW Universe, at a small 
time $\epsilon$ after 
a creation event, is not described by the averaged Einstein equations but as
a freely coasting Milne Universe. 

	Interestingly, a 
universe born as a Milne model provides just the right initial condition 
required to sort out the cosmological constant problem. It is straight
forward to formulate an action principle for gravity where the determinant
is not a dynamical quantity. Trace of the stress tensor of any matter 
field does not contribute to the dynamics of gravitation \cite{wein}.
Although this sorts out the naturalness problem of the cosmological constant,
an effective cosmological constant appears as an integration constant in this
formulation. What is needed is some physical reason that demands a flat 
space-time
solution to describe cosmology at any instant of time and our conjecture 
does precisely that.

	The following section reviews the concordance of linear 
evolution in relation to standard cosmological observations.

\vspace{.5cm}
\section
{\bf 2. A linearly coasting cosmology}
\vspace{.5cm}
\subsection
{\bf 2.1 Classical Cosmology tests}

	To our knowledge, the first exploration of concordance of a linearly
evolving scale factor with observations was conducted 
by Kolb \cite{kolb1}. Kolb obtained 
a linear evolution by a judicious choice of ``K-matter'' that makes
the universe curvature dominated at low red-shifts. At sufficiently high
red-shifts, normal matter becomes increasingly dominant. 
One could thus manage to have 
a linear coasting at low red-shifts without giving up several nice results 
of standard cosmology such as  
cosmological nucleosynthesis.  Kolb demonstrated  
that data on Galaxy number counts as a function of red-shift as well as 
data on angular diameter distance as a function of red-shift do not rule
out a linearly coasting cosmology. Unfortunately, these two tests are 
marred by effects such as galaxy mergers and galactic evolution.  For
these reasons these tests have fallen into disfavor as reliable indicators of 
a viable model.

	The variation of apparent luminosity of a ``standard candle'' 
as a function of red-shift is referred to as the Hubble test.
The discovery of Supernovae type Ia [SNe Ia] 
as reliable standard candles, raised hopes of elevating 
the status of this test to that
of a precision measurement that could determine the viability 
of a cosmological model.
The main reason for regarding these objects
as reliable standard candles are their large luminosity, small dispersion 
in their peak luminosity and a fairly accurate modeling of their 
evolutionary features. Recent measurements on 42 high red-shift 
SNe Ia's reported
in the supernovae cosmology project 
\cite{perl} together with the observations of
the 16 lower red-shift SNe Ia's of the Callan-Tollolo survey \cite{ham,ham1} 
have been used to determine the cosmological 
parameters $\Omega_{\Lambda}$ and $\Omega_{M}$ for the {\it standard model}. 
The data eliminates the 
``minimal  inflationary'' prediction defined by $\Omega_{\Lambda} = 0$ and 
$\Omega_{M} = 1$. The data can however, be used to assess a ``non-minimal 
inflationary cosmology'' defined by $\Omega_\Lambda \ne 0$,~
$\Omega_{\Lambda} +\Omega_{M} = 1$.
The maximum likelihood analysis following from such a study has yielded 
the values $\Omega_M = 0.28 \pm 0.1$ and $\Omega_\Lambda = 0.72 \pm 0.1$
\cite{perl1,perl2,wendy,branch}.

     To explore the concordance of a linear coasting cosmology,
it is convenient to consider a 
power law cosmology with the scale factor 
$a(t) = {\bar k} t^\alpha$, with ${\bar k},~\alpha$
arbitrary constants. It is straightforward to discover 
the following relation between the apparent magnitude $m(z)$, the absolute
magnitude $M$  and the red-shift $z$ of an object for such a cosmology:
\begin{equation}
m(z) = {\mathcal{M}} + 5logH_{o} + 5 log({\alpha \over H_o})^{\alpha}(1 + z) 
{\bar k}\emph{S}[{1 \over {(1 -\alpha){\bar k}}}({\alpha \over H_o})^{1 - \alpha}
( 1 - (1 + z)^{ 1- {1 \over \alpha}})] 
\end{equation}
Here $\emph{S}[X] = X, {\rm Sin}(X)$ or ${\rm Sinh}(X)$ for
$K = 0, \pm 1$ respectively, and $\mathcal{M} = M - 5log(H_o) + 25$. 
The best fit  turns out to be $\alpha = 1.001 \pm.0043$, $K = -1$.
\cite{abha}. The minimum $\chi^2$ per
degree of freedom turns out to be  1.18. 
This is comparable to the corresponding value
1.17 reported by Perlmutter et al for non-minimal inflationary 
cosmology parameter estimations. Linear coasting is as accommodating even 
for the largest red-shift supernova [1997ff] as the standard non-minimal
inflationary model.   
The concordance of linear coasting with SNe1a data
finds a passing mention in the analysis of Perlmutter \cite{perl}
who noted that the curve for $\Omega_\Lambda = \Omega_M = 0$ (for which
the scale factor would have a linear evolution) is 
``practically identical to 
$\bf{best fit}$ plot for an unconstrained cosmology''.

      The age estimate of the ($a(t) \propto t$) 
universe, deduced from a measurement of the Hubble parameter, is given
by $t_o = (H_o)^{-1}$.  The low red-shift SNe1a data \cite{ham,ham1}
gives the best value of $65~{\rm km~ sec^{-1}~ Mpc^{-1}}$ 
for the Hubble parameter.
The age of the universe turns out to be $15\times 10^9$ years.
This is $\approx$ 50\% greater than the age 
inferred from the same measurement in standard (cold) dark matter dominated
cosmology (without the cosmological constant). Such an age estimate is 
comfortably concordant with age estimates of old clusters. 

	A study of consistency of linear coasting with gravitational 
lensing statistics has recently been reported \cite{abbhh}. The expected 
frequency of multiple image lensing events is a sensitive probe for the 
viability of a given cosmology. A sample of 867 high luminosity 
optical quasars projected in a power law FRW cosmology gives an expected 
number of five lensed quasars for a power $\alpha = 1.09\pm 0.3$. This indeed
matches observations. Thus a strictly linear evolution of the scale factor
is comfortably concordant with gravitational lensing statistics.

\vspace{.2cm}
\subsection
{\bf 2.2 ``The precision'' tests}
\vspace{.2cm}

	{\bf a) The Nucleosynthesis Constraint}
	 What makes linear coasting particularly appealing is a recent
demonstration of  primordial nucleosynthesis 
not to be an impediment for a linear coasting cosmology 
\cite{annu,annu2,steig}. 
A linear evolution of the scale factor may be expected to radically effect
nucleosynthesis in the early universe. Surprisingly, 
the following scenario goes through.
 
	Energy conservation, in a 
period  where the baryon entropy ratio does not change, enables 
the distribution of photons to be described 
by an effective temperature $T$ that scales
as $a(t)T = $ constant. 
With the age of the universe estimated from the Hubble parameter being
$\approx 1.5\times 10^{10}$
years, and $T_0 \approx 2.7K$, one concludes that the age of the
universe at $T \approx 10^{10}K$ would be some four
years [rather than a few seconds as in standard cosmology]. 
The universe would
take some $10^3$ years to cool to $10^7K$. 
With such time periods being large 
in comparison to the free neutron life time, one would hardly expect any 
neutrons to survive. However, with such a low rate of 
expansion, weak interactions remain in equilibrium for
temperatures as low as $10^8K$.
The neutron - pro-ton ratio keeps falling as 
$n/p \approx exp[-15/T_9]$. Here $T_9$ is the temperature
 in units of $10^9$K and the factor of 15 comes from the n-p mass 
difference in these units. There would again
hardly be any neutrons left if nucleosynthesis were to commence at (say)
$T_9 \approx 1$. 
However, as weak interactions are still in equilibrium, once
nucleosynthesis commences, inverse beta decay would replenish neutrons by
converting 
protons into neutrons and pumping them into the nucleosynthesis channel.
With beta decay in equilibrium, the baryon entropy ratio determines a low
enough nucleosynthesis rate that can remove neutrons out of the 
equilibrium buffer at a rate smaller than the relaxation time of the 
buffer. This ensures that neutron value remains unchanged as heavier nuclei
build up. It turns out that for baryon entropy ratio 
$\eta\approx 5\times 10^{-9}$, there would just
be enough neutrons produced, after nucleosynthesis 
commences, to give $\approx 23.9\% $ Helium and 
metallicity some $10^8$ times the metallicity produced in the early
universe in the standard scenario. This metallicity is 
of the same order of magnitude as seen in lowest metallicity objects.

	The only problem that one has to contend with is 
the significantly low yields of deuterium in such a cosmology. Though deuterium 
can be produced by spallation processes later in the history of the universe,
it is difficult to produce the right amount without a simultaneous over 
production of Lithium \cite{eps}  However,
as pointed out in \cite{annu}, the amount of Helium produced is quite 
sensitive to $\eta$ in such models. In an inhomogeneous universe, therefore,
one can have the helium to hydrogen ratio to have a large variation. 
Deuterium can be produced by a spallation process much later in the history
of the universe. If one considers spallation of a helium 
deficient cloud onto a helium rich cloud, it is easy to produce deuterium
as demonstrated by Epstein \cite{eps} - without overproduction of Lithium.

	Interestingly, the baryon entropy ratio required for the right 
amount of helium corresponds to $\Omega_b \approx 0.2 $. Here $\Omega_b$ is
the ratio of the baryon density to a ``density parameter'' determined by the
Hubble constant: $\Omega_b \equiv \rho_b/\rho_c = 8\pi G \rho_b/3H_o^2$.
$\Omega_b \approx 0.2$ 
closes dynamic mass estimates of large galaxies and clusters
[see eg \cite{peebls,paddy}]. 
In standard cosmology this closure is sought to be 
achieved by taking recourse to non-baryonic cold dark matter. Thus 
in a linearly scaling cosmology, there would be no need of non-baryonic cold 
dark matter at all.

\vspace{.2cm}
{\bf b) The recombination epoch }
\vspace{.2cm}

	We describe this in some detail as the peculiarities
of the recombination epoch in
a linearly coasting cosmology are not covered in any standard (curvature
dominated) cosmology description. 

	Salient features of a linear coasting cosmology at the recombination 
epoch can be deduced by making a simplifying assumption of thermodynamic 
equilibrium just before recombination. As in standard 
cosmology, a recombination process 
that directly produces a Hydrogen atom in the ground state releases a 
photon with energy $B = 13.6 eV$ in each recombination. $n_\gamma(B)$,
the number density of photons in the background radiation with energy $B$, 
is given by [see eg. \cite{seager,paddy}]:
$$
{{n_\gamma (B)}\over n} = {{16\pi}\over n}T^3exp({{-B}\over T})
\approx {{3\times 10^7}\over {\Omega_Bh^2}}exp({-{13.6}\over \tau})\eqno{(3)}
$$
Where $\tau$ is the temperature in units of eV.
This ratio is unity at $\tau \approx .8 $ for $\Omega_Bh^2 \approx 1$
 and decreases rapidly at lower 
temperatures. Any 13.6 eV photons released due to recombination 
have a high probability of ionizing neutral atoms formed a little earlier. 
[In the following, we shall quote all results by our favored values
$\Omega_b \approx 0.2$ and the Hubble parameter 65 km/sec/Mpc]   
This process is therefore not very effective for producing a net number of 
neutral atoms. The dominant recombination process proceeds through an
excited state: $(e + p \longrightarrow H^* + \gamma; ~ H^* \longrightarrow
H + \gamma_2)$. This produces two photons, each having lesser energy than
the ionization potential of the hydrogen atom. The 2p and 2s levels provide 
the most rapid route for recombination. The 2p decay produces a single 
photon, while the decay from the 2s is by two photons. As the reverse process
does occur at the same rate, this is a non-equilibrium recombination that
proceeds at a much slower rate.
The thermally averaged cross section for the process of
recombination $(p + e \leftrightarrow H + \gamma)$ is given by 
\cite{peebls,paddy}:
$$
{{<\sigma v>}\over c} \approx 4.7\times 10^{-24}({T\over {1eV}})^{1/2}~~cm^2
\eqno{(4)}
$$
This gives the reaction rate:
$$
\Gamma = n_p<\sigma v> = 2.374\times 10^{-10}\tau^{7/4}exp(-6.8/\tau)
(\Omega_bh^2)^{1/2}~cm^{-1}\eqno{(5)}
$$
This is to be compared to the Hubble expansion rate at that
epoch, $H = H_0(T/T_0)$. Given 
the Hubble constant $(H_0 = 100h~{\rm km/sec/MPc})$ and CMB effective 
temperature 
$T_0 = 2.73 K$ now, the Hubble parameter at any temperature
turns out to be: $H = 4.7\times 10^{-25}h\tau~cm^{-1}$. This equals 
$\Gamma$ at 
$$
\tau^{-3/4}exp(6.8/\tau) = 1.96\times 10^{15}(\Omega_b)^{1/2}\eqno{(6)}
$$
A straightforward iteration gives:
$$
\tau^{-1} \approx 5.17 - 0.11 ln(\tau^{-1}) + .074 ln(\Omega_b)\approx 
(.2 )^{-1}\eqno{(7)}
$$
corresponding to a redshift given by:
$$
1 + z \approx 874.5[1 + .015 ln(\Omega_b)]^{-1} \eqno{(8)}
$$
The residual fraction of electrons turns out to be \cite{paddy}:
$$
x_e \approx ({\pi\over {4\xi (3)\sqrt{2}}})^{1\over 2}\eta^{-{1\over 2}}
({T\over m_e})^{-{3\over 4}}exp(-{6.8\over \tau})\eqno{(9)}
$$
From eqn.(6), we have
$$
x_e \approx 7.9\times 10^{-9}{{\tau^{-3/2}}\over {\Omega_b h}}\eqno{(10)}
$$
For the red-shift range $800 < z < 1200$, the approximate fractional
ionization is:
$$
x_e = {{2.4\times 10^{-3}}\over {\Omega_bh^2}}({z\over {1000}})^{12.75}
\eqno{(11)}
$$
After decoupling at $\tau = .2$, this gives a residual ionization:
$$
x_{e,res} \approx 9\times 10^{-8}(\Omega_b h)^{-1}\eqno{(12)}
$$
The only process that may still be effective at such low temperatures is
the Thompson scattering with a cross section 
$\sigma_T = 6.7\times 10^{-25}~cm^2$. The optical depth for photons
would be:
$$
\tau_\gamma = \int_0^t n_b(t)x_e(t)\sigma_T dt
= -\int_0^zn_b(z)x_e(z)\sigma_T({{dt}\over {dz}})dz\eqno{(13)}
$$
With $n_b(z) = \eta n_\gamma(z) = \eta\times 421.8(1+z)^3 ~cm^{-3}$, and
$$
{{dt}\over {dz}} = - {1\over {H_0(1+z)^2}}\eqno{(14)}
$$
one can find the red-shift at which the optical depth goes to unity. 

	If one considers the residual ionization $x_{e,res}$, we get
$$
\tau_\gamma = 4.7\times 10^{-2}\times ({z\over {1000}})^2\eqno{(15)}
$$
From this optical depth, we can compute the probability that a photon was
last scattered in the interval $(z,z + dz)$. This is given by:
$$
P(z) = e^{-\tau_\gamma}{{d\tau_\gamma}\over {dz}} \approx .94\times 10^{-5}
({z\over {1000}})exp[-0.047({z\over {1000}})^2]\eqno{(16)}
$$
$\tau_\gamma$ becomes unity at $z \approx 4610$. This implies that the 
{\it residual 
ionization} has insufficient optical depth to scatter photons from the 
decoupling epoch. From the expression for fractional ionization eqn(11),
the optical depth of the last scattering surface can be deduced to be:
$$
\tau_\gamma = 170\times ({z\over {1000}})^{14.75}\eqno{(17)}
$$
This gives:
$$
P(z) \approx 2.5({z\over {1000}})^{13.75}exp[-170({z\over {1000}})^{14.75}]
\eqno{(18)}
$$
$\tau_\gamma$ goes to unity at $z_R \approx 703$. This $P(z)$ can be approximated by 
a Gaussian centered at $z_R \approx 703$ with a width $\Delta z \approx 51.8$.

	An important scale that determines the nature of CMB anisotropy is 
the curvature scale which is the same as the Hubble radius for the linear
coasting. The angle subtended today, by the Hubble radius at $z_R = 703$, is
determined by
$$
{{1+z_R}\over 2}{\theta\over 2} = {\rm sinh[{{d(\theta )(1+z_R)}\over {2a_0}}}]
\eqno{(19)}
$$
Here $d(\theta) = d_H(t_R) = H(t_R)^{-1} = [H_0(1+z_R)]^{-1}$.
This gives:
$$
({{1+z_R}\over 2}){\theta\over 2} ={\rm sinh({1\over 2}})\eqno{(20)}
$$
or $\theta_H \approx 10$ minutes. 

	In standard cosmology, the {\it sound horizon} is of the same order as 
the Hubble length. The Hubble length determines the scale over which physical
processes can occur coherently. In a linear coasting, the Hubble length is 
precisely the inverse of the curvature scale. However, the sound horizon 
($s^*$) is
much larger. Strictly speaking, the particle as well as the sound horizon are
infinite for a linear coasting cosmology. For our purpose, it suffices to 
take the epoch of birth of pressure waves as the epoch of baryon production. 
We take this to be the QGP phase transition epoch $T_{QGP} \approx 10^{12}K$. 
The distance a sound wave travels from this epoch till recombination, 
would subtend an angle which can be refereed to as the {\it sound horizon angle}:
$$
\theta^* \approx {1\over {\sqrt{3}}}ln({T_i\over T_f})\times {2\over {1+z^*}}
\eqno{(21)}
$$
This is $\approx 2^o$ for $T_i = T_{QGP}$ and $T_f\approx 10^3K$ 
corresponding to $z^* \approx 705$. 
The angle subtended by the sound horizon scale is thus roughly 12 times that 
subtended by the curvature length scale of ten minutes.
The photon diffusion scale is determined by the thickness of the LSS.
With $z^* \approx 705$ and $\Delta z \approx 51$, this gives an angular
size which is roughly one fourteenth of the Hubble length at the LSS. 
This subtends an angle of 43'' at the current epoch.

	The above scales in principle determine the nature of CMB anisotropy.
The CMB effectively ceases to scatter when the optical
depth to the present drops to unity. After last scattering, the photons 
effectively free stream. On the LSS, the photon
distribution may be locally isotropic while still possessing inhomogeneities
i.e. hot and cold spots, which will be observed as anisotropies in the 
sky today [see eg. \cite{hu,tegmark}. 
As described in the Appendix, temperature fluctuations, determined by the 
potential and density perturbations, are expressible by an expansion in terms
of eigenmodes of the generalized Laplace operator $\nabla^2$ with eigenvalues 
$-k^2$.
The phase of oscillation is frozen in at last scattering. 
The critical wave number $k_A \equiv \pi/s^*$ corresponds to the
sound horizon at that time. Longer wavelengths will not have evolved
from the initial conditions and possess $\psi/3$ gravitational
potential fluctuations after
gravitational red-shift \cite{hu,tegmark}. This combination of the intrinsic 
temperature
fluctuation and the gravitational red-shift is the ``Sachs - Wolfe effect''.
Shorter wavelengths can be frozen at different phases of the ${\rm cos}(ks^*)$
oscillation for adiabatic perturbative modes and as ${\rm sin}(ks^*)$ for 
isocurvature fluctuation modes. For adiabatic modes as
a function of $k$ there will be a harmonic series of 
temperature fluctuation peaks with $k_m = mk_A = m\pi/s^*$ for the $mth$
peak. Odd peaks represent compression phase (temperature crests), whereas 
even peaks represent the rarifaction phase (temperature troughs), inside 
potential wells. In the isocurvature 
case, just as in the adiabatic case, the self gravity of the photon baryon 
fluid essentially drives the oscillations. Unlike the adiabatic case, it is
the sine rather than the cosine oscillations that are driven now. Peaks occur
at $k = (m-1/2)k_A$ with all even peaks being enhanced by the baryon drag.
 More exotic models might produce a phase shift leading to a fluctuation
${\rm cos}(ks^* + \phi)$. This would shift the location of the first peak 
while 
leaving the spacing between the peaks the same: $k_m - k_{m-1} = k_A$. 
Thus the sound horizon
at last scattering should be measurable from the CMB.

	Subtle complications that arise in our CMB anisotropy study can be 
tackled in the same manner that deals with them in the standard model. For
example, 
in the total variance of temperature fluctuation, it can be seen that 
the photon density and potential fluctuations cancel the velocity (Doppler) 
fluctuations were the sound speed exactly $c_s = 1/\sqrt{3}$. However, for
$c_s < 1/\sqrt{3}$, the locations of the peaks for the temperature variance
coincides with those of the photon density and potential fluctuations
[see eg \cite{tegmark}]. The wave number $k = 1$, in units of the curvature
scale, would correspond to a length on the LSS that subtends an angle of 
$10'$.
It is straightforward to determine the peak location for the adiabatic and 
isocurvature perturbations for the primary SW effect. For adiabatic
modes, compression peaks occur for odd values of $m$ at angles 
$\theta^{ad}_m = 120/m\pi $ minutes. For isocurvature modes they occur at 
even m at
$\theta^{iso}_m = 120/(m-{1\over 2})\pi $ minutes. Fluctuations would have
a decreasing amplitude for smaller angles due to photon diffusion that makes 
the coupling between the baryon - photon fluid bleed for small scales as it 
vanishes at $43''$.

	All modes corresponding to angles greater than 10 minutes correspond 
to eigenmodes $0 < k < 1$. These are supercurvature modes. The location
of the largest (adiabatic) wavelength peak is $k = \pi/12 \approx 1/4$. 
As explained in the appendix \cite{lyth,wu}, the eigenfunctions of
supercurvature modes are suppressed for open models. However, for $k$ as
low as $1/4$ the suppression of the eigenfunction is merely by a factor of the
order unity. The relative amplitudes of the $k$ modes is determined by an
initial power spectrum that is set by an {\it ab initio} ansatz.  
The suppression of the supercurvature mode with $k \approx 1/4$
can be countered by a corresponding change in the initial power spectrum. 

	The exact profile of the anisotropy would be determined by the choice
of the nature of initial conditions (adiabatic or isocurvature), the chosen 
initial power spectrum, and the growth of perturbations after $z^*$ 
(decoupling). These determine the late or the {\it integrated SW effect}, 
aspects of
reionization etc.

	The main point we make in this article is that in spite
of a significantly different evolution, the recombination history of a 
linearly coasting cosmology gives the location of peaks for the primary
acoustic peaks in the 
same range of angles as that given in Standard Cosmology. 
Given that none of the alternative anisotropy formation scenarios 
provide a compelling {\it ab initio} model  \cite{hu1995} , 
it is perhaps best to keep an open 
mind to all possibilities. As the large scale structure and CMB anisotropy 
data continue to accumulate, one could explore the general principles 
for an open coasting cosmology to aid in the empirical 
reconstruction of a consistent model for structure formation.

	Finally, we are tempted to mention that a linear coasting cosmology
presents itself as a falsifiable model. It is encouraging to observe its
concordance !! In standard cosmology, falsifiability has taken on a backstage
- one just constrains the values of cosmological parameters subjecting 
the data to Bayesian statistics.   

\vspace{.2cm}
{\bf Acknowledgments:} 
\vspace{.2cm}

	The authors thank G. F. R. Ellis and J. Ehlers for clarifying
aspects of the averaging problem. Fruitful discussions with T. Saurodeep and S. Engineer are gratefully
acknowledged.

\vspace{.2cm}
{\bf Appendix:}
\vspace{.2cm}
	Subsequent to decoupling, perturbations of the last scattering 
surface [LSS] and the intervening space, leave an imprint on the 
streaming microwave background photons observed at the present epoch. To
describe the 
{\it gross} features of perturbations of the model we start by 
writing the background line element as
$$
ds^2 =~ ^{(0)}g_{\mu\nu}(x)dx^\mu dx^\nu = dt^2 - a^2(t)\gamma_{ij}dx^idx^j
= a^2(\eta)(d\eta^2 - \gamma_{ij}dx^idx^j)\eqno{(A.1)}
$$
where $\eta$ is the conformal time $d\eta \equiv a^{-1}dt$.
$$
\gamma_{ij} = \delta_{ij}[1+ {1\over 4}K(x^2 + y^2 +z^2)]^{-2}\eqno{(A.2)}
$$
where $K = -1$ for the $\eta = $ constant hypersurface describing an 
open model's space-like section.

	Assuming the perturbations to be described by the perturbed 
Einstein Equations: $\delta G_{\mu\nu} = \delta T_{\mu\nu}$, the metric
can be expanded as usual in terms of the scalar, vector and tensor modes
[see eg. \cite{brand}]. The gauge invariant scalar perturbation equations are:
$$
\nabla^2\Phi -3H\phi ' -3(H^2 - K)\Phi = 4\pi Ga^2\delta\epsilon^{gi}
\eqno{(A.3a)}
$$
$$
(a\Phi)'_{,i} = 4\pi Ga^2(\epsilon_o + p_o)\delta u_i^{gi}\eqno{(A.3b)}
$$
$$
\Phi'' + 3H\Phi' + (2H' + H^2 - K)\Phi = 4\pi Ga^2\delta p^{gi}\eqno{(A.3c)}
$$
Here, $\nabla^2\Phi \equiv \gamma^{i,j}\Phi_{;i;j}$, is the wave operator
for the open model. $H \equiv a'/a$, where $'$ is a derivative with respect 
to conformal time, and finally the $\delta\epsilon^{gi},~\delta u_i^{gi}$ and
$\delta p^{gi}$ are the gauge invariant density, velocity and pressure 
parameters respectively \cite{brand}. 
These equations are valid whenever linear perturbation theory is valid. This
requires $|\Phi | << 1$ but not necessarily 
$|\delta\epsilon/\epsilon| << 1$. 
The above equations combine to give:
$$
\Phi '' + 3H(1+c_s^2)\Phi ' - c_s^2\nabla^2\Phi
+ [2H' + (1+3c_s^2)(H^2 - K)]\Phi = 4\pi Ga^2\tau\delta S\eqno{(A.4)}
$$
Here the parameters $c_s, ~\tau$ are determined in terms of the matter,
radiation and entropy densities $\epsilon_m,~\epsilon_\gamma,~S$ 
and are given by:
$$
c_s^2 = {1\over 3}(1 + {3\over 4}{\epsilon_m\over \epsilon_\gamma})^{-1},
~~~ \tau = {{c_s^2\epsilon_m}\over S}\eqno{(A.5)}
$$ 
Entropy perturbations, $\delta S$, also called isocurvature perturbations,
can be generated if the different matter components
are distributed non-uniformly in space but with uniform total energy density
and hence uniform curvature at the beginning.

	For a radiation dominated epoch, the evolution of adiabatic 
perturbations ($\delta S = 0$)is given by putting 
$c_s \approx 1/\sqrt{3}$ when eqn(A.4) reduces to:
$$
\Phi'' + 4\Phi' + {k^2\over 3}\Phi + 4\Phi = 0\eqno{(A.6)}
$$
where we define $-k^2$ as the eigenvalue for $\nabla^2$.
A straightforward solution to this equation is: 
$\Phi \longrightarrow t^{-2}exp(ik\eta/\sqrt{3})$. This form for $\Phi$,
together with eqn(A.3a) determine the density perturbations in the
radiation dominated epoch provided we
have an ansatz for an initial power spectrum. It is also straightforward to
solve the potential equations in the matter dominated epoch as well.

In general [see eg \cite{hu}] it is convenient to expand
cosmological perturbations in a series of eigenfunctions
of the Laplacian. Firstly, each mode (each term in 
the series) evolves independently with time. This makes is easy to evolve
a given initial perturbation forward in time. Secondly, by assigning a 
Gaussian probability distribution to the amplitude of each mode, one can 
generate a homogeneous Gaussian random field. Such a field consists of an
ensemble of possible perturbations. It is supposed that the perturbations
seen in the observable universe is a typical member of the ensemble. 
The schotastic properties of a Gaussian random field are determined by its 
two point correlation function $<f(1)f(2)>$, where $f$ is the perturbation 
and the brackets denote the ensemble average. For a homogeneous field, the
correlation depends only on the distance between the two points.

For the expansion of perturbations in terms of the Laplacian 
with eigenvalues $-k/a^2$, modes with real $k^2 > 1$ 
provide a complete orthonormal 
basis for $L^2$ functions \cite{yadlom,lyth}.
They vary appreciably on scales less than the curvature scale $a$ and
are called subcurvature modes. A related wave number and a related radial
coordinate are defined as:
$$
q^2 \equiv k^2 - 1, ~~ \chi \equiv{\rm sinh^{-1}r} 
$$
A typical expansion of the wave mode is:
$$
f(\chi,\theta,\phi,t) = \int_0^\infty dq\sum_{lm}f_{klm}(t)
Z_{klm}(\chi,\theta,\phi)\eqno{(A.7)}
$$
Where $Z_{klm} \equiv \Pi_{kl}(\chi)Y_{lm}(\theta,\phi)$, and the
radial functions are:
$$
\Pi_{kl} = {{\Gamma(l + 1 + iq)}\over {\Gamma(iq)}}{1\over {\sqrt{\rm
{sinh\chi}}}}P_{iq - 1/2}^{-l-1/2}({\rm cosh}\chi)\eqno{(A.8)}
$$
normalized as:
$$
\int_0^\infty \Pi_{kl}(\chi)\Pi_{k'l'}(\chi){\rm sinh}^2\chi d\chi = 
\delta(q-q')\delta_{ll'}
$$
$$
\int Z^*_{klm}Z_{k'l'm'}d{\cal V} = \delta(q-q')\delta_{ll'}\delta_{mm'}
\eqno{(A.9)}
$$
The constant non-zero phase of $\Pi_{kl}$ can be dropped by defining
the real function:
$$
\Pi_{kl} \equiv N_{kl}{\hat \Pi_{kl}}
$$
$$
{\hat \Pi_{kl}} \equiv q^{-2}({\rm sinh}\chi)^l({{-1}\over {\rm{sinh}\chi}}
{d\over {d\chi}})^{l+1}{\rm cos}(q\chi)
$$
$$
N_{kl} \equiv \sqrt{{2\over pi}} q^2[\Pi_{n=0}^l(n^2+q^2)]^{-1/2}\eqno{(A.10)}
$$

The problems with these modes is 
that they are inadequate to describe 
perturbations over scales larger than the curvature scale.
For this purpose, while considering 
perturbations in an open universe, one should
retain not only the subcurvature modes (defined as eigenfunctions of the
Laplacian with eigenvalues less than -1 in units of curvature scale), but 
also the supercurvature modes whose eigenvalues lie between 0 and -1. 
All modes
must be included to generate the most general homogeneous Gaussian random
field even though they may not be linearly independent. The reason for this
is the following:

  With	cosmological perturbations assumed to be Gaussian in the 
regime of linear evolution, a Gaussian perturbation is defined as one whose
probability distribution functions are multivariate Gaussians and its 
stochastic properties are completely determined by its correlation function.
The perturbation turns out to be homogeneous with 
the correlation function depending only on 
the distance between the points. 

	If one merely includes the subcurvature modes, it is easy to 
deduce the form for the correlation function \cite{lyth,yadlom}:
$$
\xi_f = \int_1^\infty {{dk}\over k}P_f(k){\rm{{ sin(qr)}\over {q~sinhr}}}
\eqno{(A.11)}
$$
Setting $r = 0$ gives the mean square value:
$$
\xi_f(0) \equiv <f^2> = \int_1^\infty {{dk}\over k}P_f(k)\eqno{(A.12)}
$$
Therefore, by expanding a
perturbation in terms of subcurvature modes, the correlation is bounded by:
$$
{{\xi_f(r)}\over {\xi_f(0)}} < {\rm {r\over {sinhr}}}\eqno{(A.13)}
$$
$q\longrightarrow 0$ does not correspond to infinitely large scales, but to 
scales of the order of the curvature scale.

	Thus including only the subcurvature modes generates a Gaussian 
perturbation whose correlation function necessarily falls off faster than
${\rm r/sinhr}$. This reflects the fact that each supercurvature mode varies 
strongly on a scale no bigger than the curvature scale. A random superposition
of such modes will hardly ever be nearly constant on a scale much bigger than
the curvature scale. This is precisely what the lack of correlation on large 
scales tells us.

  One could consider correlation on 
arbitrarily large scales  by including the 
super curvature modes. For $-1 < q^2 < 0$ the analytic continuation of the 
radial function $\Pi_{kl}$ gives the supercurvature modes:
$$
\Pi_{kl} \equiv N_{kl}{\hat \Pi_{kl}}
$$
$$
{\hat \Pi_{kl}} \equiv |q|^{-2}({\rm sinhr)^l({{-1}\over {sinhr}}
{d\over {dr}})^{l+1}cosh(|q|r)}
$$
$$
N_{k0} \equiv \sqrt{{2\over \pi}} |q|
$$
$$
N_{kl} \equiv \sqrt{{2\over \pi}} |q|[\Pi_{n=1}^l(n^2+q^2)]^{-1/2}
~~~~(l > 0) \eqno{(A.14)}
$$

These supercurvature modes go as $exp[-(1 - |q|)r]$ at large r. With the
volume element 
$d{\cal V} = {\rm sinh^2r~ sin\theta~ d\theta~ dr~ d\phi}$ the integral over
all of space of a product of any two of them diverges. The modes are therefore
not orthogonal let alone orthonormal. In a finite region of space they are not
linearly independent of the subcurvature eigenfunctions. None of this matters
for the purpose of generating a Gaussian perturbation. The supercurvature 
modes add to the expansion (A.7), an additional:
$$
f^{SC}(r,\theta,\phi,t) = \int_0^1 d(iq)
\sum_{lm}f_{klm}(t)Z_{klm}(r,\theta,\phi)\eqno{(A.15)}
$$
From this, the supercurvature contribution to the correlation
function is seen to be \cite{lyth}:
$$
\xi^{SC}_f(r) = \int_0^1 {{dk}\over k}P_f(k){{{\rm sinh(|q|r)}\over 
{{\rm |q|sinhr}}}}
\eqno{(A.16)}
$$

	Consider a supercurvature mode corresponding to a peak
at $k \approx 1/3$ or $q = 2\sqrt{2} i/3$ in units of curvature scale. For
such a mode, the correlation function is suppressed by a factor 
${\rm sinh(|q|r)/(|q|sinh(r))} \approx 2/3$. This is a suppression by a factor
of the order unity and can be compensated by an appropriate initial power
spectrum.

The spectrum of initial fluctuations can be characterized by a power law 
$|\delta_k|^2 = VAk^n$ where $n$ is a spectral index and $A$ is the amplitude
at very early epochs. The values of these parameters should emerge from the
physical model which describes the the production of the initial spectrum.
In the absence of any reliable theoretical prediction for $A$ and $n$, it
is best to treat them as free parameters which can be determined by 
comparison with observations.

\vspace{1cm}

\bibliography{plain}

\end{document}